\documentclass[iop]{emulateapj}       
\usepackage{multirow}
\usepackage{bm}
\usepackage{amssymb}
\usepackage{amsmath}
\usepackage{latexsym}
\usepackage{graphicx}
\usepackage{ifsym}
\usepackage{bm}
\usepackage{tabularx}
\usepackage{hyperref}
\usepackage{enumitem}

\shorttitle{MagAO H$\alpha$ Imaging of $\eta$ Car}
\shortauthors{Wu et al.}

\slugcomment{}

\begin{document}
\title{{\textbf {\Large R\lowercase{esolving the} H$\alpha$\lowercase{-emitting} R\lowercase{egion in the} W\lowercase{ind of} $\eta$ C\lowercase{arinae}}}}


\author{Ya-Lin Wu, Nathan Smith, Laird M. Close, Jared R. Males, and
  Katie M. Morzinski}

\affil{Steward Observatory, University of Arizona, Tucson, AZ 85721,
  USA; yalinwu@email.arizona.edu\\{\it Published in ApJL}}

\begin{abstract}
  The massive evolved star $\eta$ Carinae is the most luminous star in
  the Milky Way and has the highest steady wind mass-loss rate of any
  known star.  Radiative transfer models of the spectrum by Hillier et
  al.\ predict that H$\alpha$ is mostly emitted in regions of the wind
  at radii of 6--60 AU from the star (2.5--25 mas at 2.35 kpc). We
  present diffraction-limited images (FWHM $\sim$ 25 mas) with Magellan
  adaptive optics (MagAO) in two epochs, showing that $\eta$ Carinae
  consistently appears $\sim$2.5--3 mas wider in H$\alpha$ emission
  compared to the adjacent 643 nm continuum. This implies that the
  H$\alpha$ line-forming region may have a characteristic emitting
  radius of 12 mas or $\sim$30 AU, in very good agreement with the
  Hillier stellar-wind model.  This provides direct confirmation that
  the physical wind parameters of that model are roughly correct,
  including the mass-loss rate of $\dot{M}$ = 10$^{-3}$ $M_{\odot}$
  yr$^{-1}$, plus the clumping factor, and the terminal
  velocity. Comparison of the H$\alpha$ images (ellipticity and PA) to
  the continuum images reveals no significant asymmetries at
  H$\alpha$. Hence, any asymmetry induced by a companion or by the
  primary's rotation do not strongly influence the global H$\alpha$
  emission in the outer wind.
\end{abstract}


\section{Introduction}  

The strong mass loss of $\eta$ Carinae provides important clues for
our understanding of the late evolutionary phases of very massive
stars. Classified as a luminous blue variable (LBV), $\eta$ Carinae
has exhibited dramatic instability and suffers significant mass loss
evidenced by its historical eruptions and its strong present-day
wind. Over the past 800 years, repeating eruptions have created the
spectacular Homunculus Nebula and numerous outer ejecta (e.g.,
\citealt{morse01,S03b,K16}) that have removed 20~$M_\sun$ or more from
the star. Despite this extreme mass ejection, the star apparently
survived the 19th century Great Eruption \citep{sf11}, and its
present-day wind has the highest mass-loss rate of any hot star. While
the past eruptive mass loss is so strong that it must have been a
continuum-driven super-Eddington wind or a non-terminal explosion
\citep{S03b,owocki04,so06,smith08,os16}, its present-day wind is near
the limit of what can be powered by conventional line-driven winds.
As such, the strong wind of $\eta$ Carinae is an important test case
for understanding the physics of strong stellar-wind mass loss from
hot stars.

LBVs and related hot supergiants have such dense ionized winds that
their H$\alpha$ emission is extremely strong, and comes from radii
that can be vastly more extended than the star's surface. The
H$\alpha$-emitting region of the famous LBV star P~Cygni, which is
located closer to us than $\eta$~Car, has been resolved by optical
interferometry \citep{vakili97}. It has an apparent size of $\sim$5
mas in H$\alpha$, which is about 20 times larger than the
visible-wavelength continuum photosphere. Some faint extended
structure around P Cygni seen in adaptive optics imaging in H$\alpha$
was interpreted as clumpy structures in the outer wind or inner nebula
\citep{chesneau00}.

Based on a comparison with the UV to near-IR spectrum of the central
star observed by the {\it Hubble Space Telescope} ({\it HST}),
\cite{H01} constructed a 1D radiative transfer model of $\eta$ Car's
spectrum, indicating a very high present-day mass-loss rate of
10$^{-3}$ $M_{\odot}$ yr$^{-1}$.  This model indicates that the
expected H$\alpha$-emitting region is extended, at roughly 20--200
stellar radii. This translates to about 1200--12000~$R_{\odot}$ for
the adopted stellar radius of 60 $R_{\odot}$, or radii of 6--60 AU
(2.5--25 mas at $D$ = 2.35 kpc; \citealt{S06}).  Directly measuring
the extent of this H$\alpha$ emission provides a way to test models of
the wind mass loss.

For $\eta$ Car, the H$\alpha$-emitting region of the wind has not yet
been directly resolved. This is partly because $\eta$ Car is a more
difficult target than P~Cygni, due to its larger distance and because
it is surrounded by very complex nebulosity, which is at least partly
attributed to a series of shells and clumps produced by the
interacting wind shocks in this massive binary system.  Imaging and
long-slit spectroscopy with {\it HST} have not been able to directly
resolve the stellar wind's H$\alpha$ emission, although they have shown
complex and time-variable extended structures in the nebula within
0\farcs1--0\farcs3 of the star
\citep{morse01,smith04,gull09,madura12}.  At ultraviolet (UV)
wavelengths, the wind of $\eta$ Car has a more extended halo because
of bound--bound scattering opacity and possibly because of obscuration
of the central star's continuum.  This extended UV wind emission, seen
mainly in Fe~{\sc ii} and [Fe~{\sc ii}] line emission, has been
directly resolved with {\it HST} at radii of 30--200 mas from the star
\citep{hillier06}.  The outer wind has also been resolved with radio
interferometry with a size of $\sim$1\arcsec, due to the very large
extent and high optical depth of free-free emission
\citep{duncan99,dw03}. The radio emission is highly variable and is
strongly influenced by time-dependent ionization of circumstellar
material from the hot companion star.  The highest angular resolution
for $\eta$~Car has been achieved with near-IR interferometry, which
probes deep enough into the wind to see significant asymmetry at
angular scales of 5 mas \citep{vanboekel03,groh10,W16}.  The detected
asymmetry has been interpreted as evidence for a bipolar primary wind
geometry induced by rotation, as well as strong asymmetry in the
equatorial emitting regions due to the shock cone structure in the
colliding winds of the binary system.

Mass-loss rates derived from these studies vary from 0.5--1.6 $\times$
10$^{-3}$ $M_\sun$ yr$^{-1}$.  The high end of these values was
derived from emission in the polar wind \citep{vanboekel03}.  Lower
values of 0.5 $\times$ 10$^{-3}$ $M_{\odot}$ yr$^{-1}$ for the primary
star's mass-loss rate are derived from models of the X-ray emission
from the colliding wind binary \citep{parkin09}. Since the model of
\citet{H01} is based on a 1D code that adopts spherical symmetry, the
mass-loss rate may vary with latitude and the integrated spectrum on
which the model is based may be a representative average.  Indeed,
analysis of line profiles of H$\alpha$ emission reflected by the dusty
Homunculus Nebula suggests that the velocity and density of the
present-day stellar wind do indeed vary from equator to pole
\citep{S03a}, with higher density and faster outflow toward the poles
(the poles of the wind are oriented along the same direction as the
polar axis of the nebula).  From the reflected line profiles, the
emission components do not show a strong latitude dependence, while
blueshifted P Cygni absorption, tracing a single line of sight to the
star's photosphere, shows significant latitude dependence that points
toward a rotationally shaped wind \citep{S03a}. Allowing the wind to
be aspherical introduces a number of free parameters into the model
\citep{hillier06}, allowing for direct imaging of the H$\alpha$ region
to provide helpful constraints.

To directly measure the location of the H$\alpha$ line-forming region
of the wind would require a $<$40 mas resolution at optical
wavelengths. Here, we report diffraction-limited imaging of $\eta$ Car
using the Magellan adaptive optics (MagAO; \citealt{C12,C13}; \citealt{Males14}; \citealt{M14}) in 2016 and 2017. We
show that $\eta$ Car appears $\sim$2.5--3 mas more extended in
H$\alpha$ than the 643 nm continuum, thereby constraining the
dimensions of the H$\alpha$-forming region in $\eta$ Car's wind.

\begin{figure*}
\center
\includegraphics[angle=0,width=\linewidth]{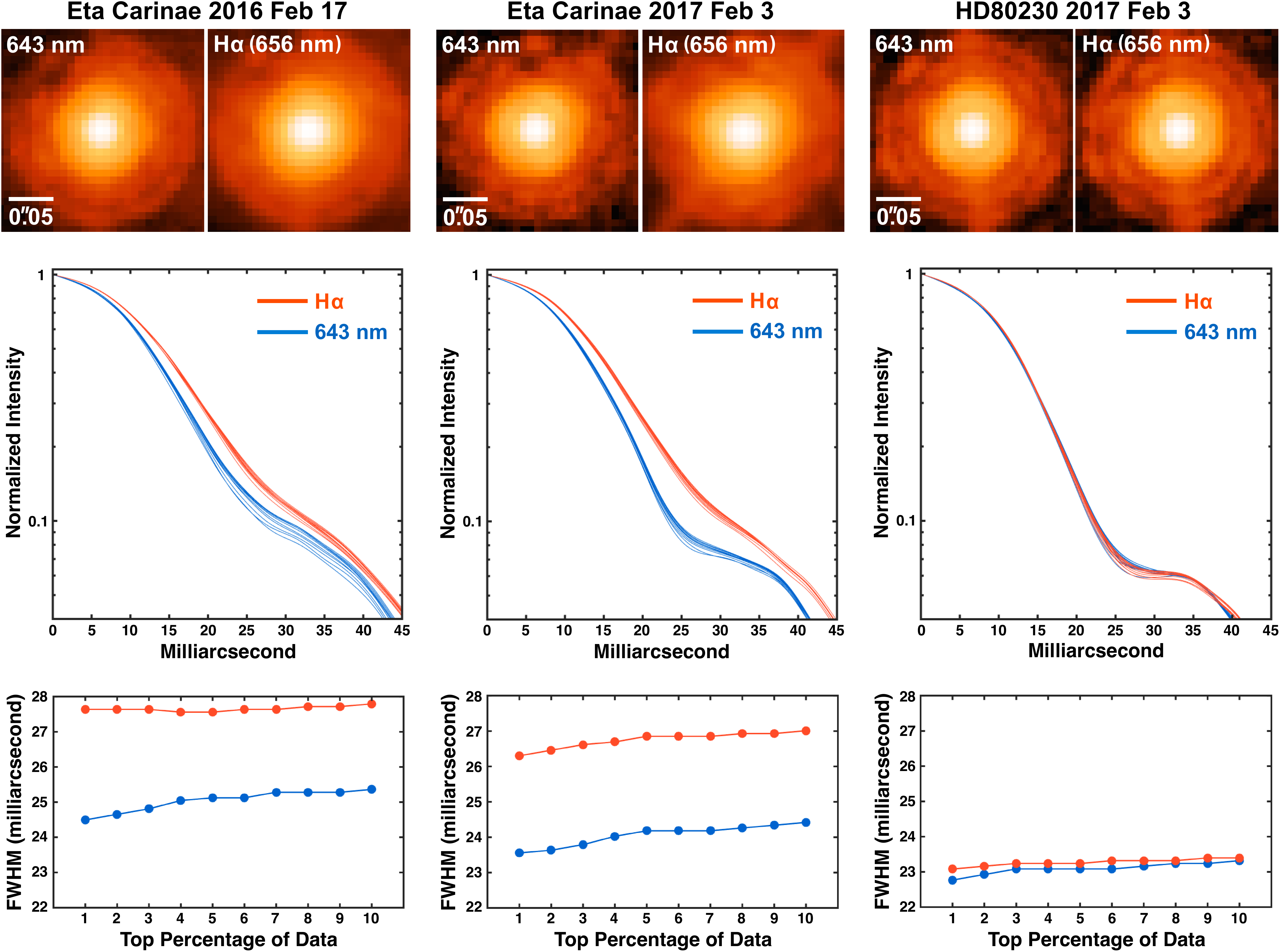}
\caption{Radial profiles and FWHMs of $\eta$ Car and the PSF reference
  star HD 80230. Top panels: H$\alpha$ and 643 nm images using the top
  1\% of the data. North is up and east is left. Middle panels:
  normalized radial profiles for images processed with the top 1\%, 2\%,
  3\%, ..., 10\% of the data. The nearly identical PSFs of HD 80230
  indicate that our observations are not just an effect of seeing
  changes: H$\alpha$ of $\eta$ Car is always resolved. Bottom panels:
  H$\alpha$ of $\eta$ Car is about 2.5 to 3 mas wider than 643 nm.}
\label{f.allobs}
\end{figure*}

\begin{figure*}
\center
\includegraphics[angle=0,width=\linewidth]{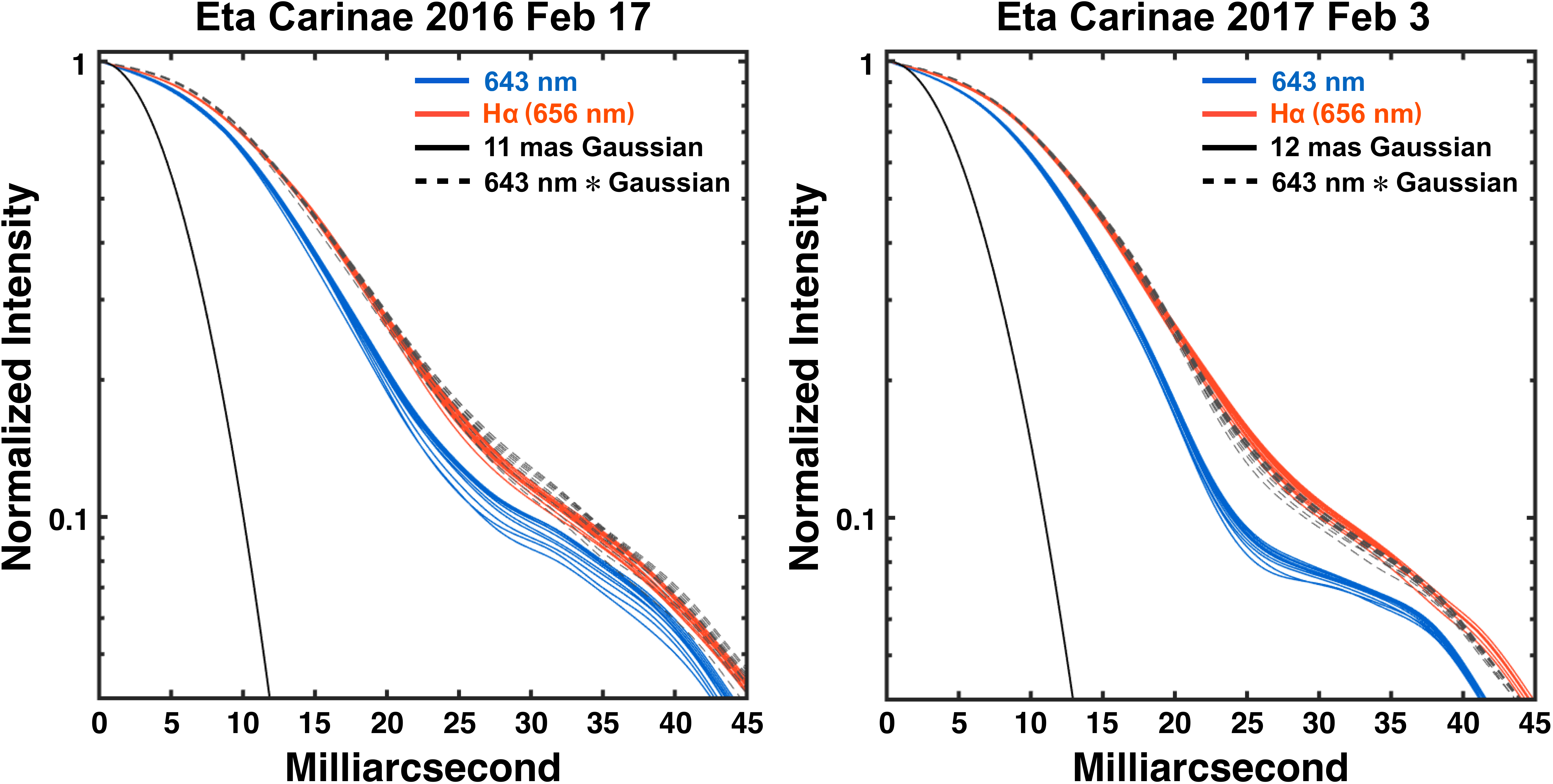}
\caption{Convolving the 643 nm profiles with an FWHM = 11 or 12 mas
  Gaussian yields a nice fit to the H$\alpha$ profiles. This implies
  that the H$\alpha$-emitting region around $\eta$ Car may have a
  characteristic size around 30 AU.}
\label{f.convolution}
\end{figure*}

\section{Observations and Data Reduction}  

We obtained images of $\eta$ Carinae on 2016 February 17, and again on
2017 February 3 with the MagAO system operated with 300 and 350 corrected modes,
respectively. These two epochs were obtained at orbital phases in the
binary system of roughly 0.28 and 0.46, about 1.5--2.5 year after
periastron.  Observations in the H$\alpha$ filter (656 nm;
$\Delta\lambda$ = 6.3 nm) and the neighboring continuum filter (643
nm; $\Delta\lambda$ = 6.1 nm) were interleaved (H$\alpha$, continuum,
H$\alpha$, continuum, etc.) so as to wash out any small effects of
variability of seeing. In our second-epoch observations, we also
imaged the nearby red giant HD~80230 ($V\sim$ 4.3 mag), which is unlikely to have any
extended H$\alpha$ emission, in the same manner to calibrate the shape
of the point spread function (PSF) on an unresolved source. Table \ref{tab:MagAO_obs}
summarizes our observations.

Raw data were dark-subtracted, de-rotated, and registered with
IRAF\footnotemark[1]\footnotetext[1]{IRAF is distributed by the
  National Optical Astronomy Observatories, which are operated by the
  Association of Universities for Research in Astronomy, Inc., under
  cooperative agreement with the National Science
  Foundation.}(\citealt{T86}, \citeyear{T93}). Image quality was
gauged by the peak intensity of the core. To achieve optimal
resolution, we only average-combined the top 1\%, 2\%, 3\%, ..., 10\% of the
reduced frames and measured the FWHM. Since the signal-to-noise
ratios (S/Ns) of our average-combined images are high (S/N $\sim$ 3000--7000), we can achieve accurate measurements of the FWHM values. Plate scale for H$\alpha$ in MagAO images is 7.85 mas pix$^{-1}$ \citep{C13}.

\begin{deluxetable*}{@{}lllcccc@{}}
\tablewidth{\linewidth}
\tablecaption{MagAO Observations \label{tab:MagAO_obs}}
\tablehead{
\colhead{Object} &
\colhead{Date} &
\colhead{Seeing} &
\colhead{Filter} &
\colhead{Speed} &
\colhead{Mode} &
\colhead{Exposure} 
}
\startdata
\multirow{4}{*}{$\eta$ Carinae}	& \multirow{2}{*}{2016 Feb 17} 	& \multirow{2}{*}{$\sim$0\farcs65}	&	643 nm	&	990 Hz	&	300	&   32 ms $\times$ 5000\\
					& 	 					& 							&	656 nm  	& 	990 Hz 	&	300	&   32 ms $\times$ 5000\\
					& \multirow{2}{*}{2017 Feb 03} 	& \multirow{2}{*}{$\sim$0\farcs5}	&	643 nm	&	1 kHz	&	350	&   32 ms $\times$ 2500\\
					& 						& 							&	656 nm  	& 	1 kHz 	&	350	&   32 ms $\times$ 2539 \\ \\
\multirow{2}{*}{HD 80230}	& \multirow{2}{*}{2017 Feb 03} 	& \multirow{2}{*}{$\sim$0\farcs75}	&	643 nm	&	1 kHz	&	350	&   32 ms $\times$ 2500	\\					&  						& 							&	656 nm  	& 	1 kHz 	&	350	&   32 ms $\times$ 2500	
\enddata
\end{deluxetable*}

\section{Results and Interpretation}

Figure \ref{f.allobs} shows our two epochs of MagAO observations of
$\eta$~Car. In the top panels, we show the images of $\eta$~Car and
the PSF reference star HD~80230 in 643 nm and H$\alpha$ from the top
1\% of the data. The first two Airy rings are clearly seen. In the
middle panels, we plot the normalized radial profiles in images
processed with the top 1\%, 2\%, 3\%, ..., 10\% of the data. It is clear that
$\eta$~Car appears more extended in H$\alpha$ than in the continuum in
both epochs, while HD~80230 has almost identical PSFs in both
filters. This demonstrates that $\eta$~Car's different profile shapes
between line and continuum are genuine, rather than from atmospheric
turbulence or imperfect AO corrections. This difference is better seen
in the bottom panels, where we show that $\eta$~Car's H$\alpha$ FWHM
is approximately 2.5--3 mas wider than its 643 nm continuum. We note
that we did not enlarge the 643 nm images by the ratio of wavelength
difference (2\%) because the slightly poorer AO performance at bluer
wavelengths might enlarge the 643 nm PSF by a similar amount, as
hinted by the nearly identical H$\alpha$ and 643 nm FWHMs of HD~80230.
In any case, the difference in profile widths between line and
continuum is much larger than the difference in wavelength.

The distribution of H$\alpha$ emission is expected to be highly
asymmetric because of the bipolar primary wind \citep{S03a}, as well
as the influence of the companion star \citep{D96}, but the
resolutions of our images are not high enough to reveal unambiguously
any weak asymmetric structure. We find that the ellipticity and PA are
indistinguishable between the H$\alpha$ and continuum images. Any real
asymmetry is evidently more compact than we can probe, at size scales
of $\sim$5 mas as seen in the near-IR interferometric results
\citep{vanboekel03,groh10,W16}. To further characterize the extent of
the H$\alpha$-forming region in $\eta$~Car's wind, in Figure
\ref{f.convolution} we show that the 2.5--3 mas ($\sim$10\%) increase
in the H$\alpha$ FWHM can be reproduced by convolving the 643 nm PSF
with an FWHM$\sim$12 mas Gaussian. This implies that the
H$\alpha$-forming region may be on the order of 12 mas in size,
corresponding to $\sim$25--30 AU at a distance of 2.35 kpc.

This characteristic scale is remarkably consistent with the
predictions in \cite{H01} that H$\alpha$ is mostly emitted at a
separation of $\sim$20--200 stellar radii, with a peak at a radius
of about 65 stellar radii (see Figure 15 in \citealt{H01}).  With the
stellar radius of 60 $R_{\odot}$ that \citet{H01} adopted for
$\eta$~Car, this translates to a range of radii from 6 to 60 AU and a
peak at a radius of $\sim$20 AU.  Our finding of the observed extent
of H$\alpha$ emission around $\eta$ Car at 25--30 AU therefore
provides direct confirmation of the globally averaged values of
$\dot{M}$ = 10$^{-3}$ $M_{\odot}$ yr$^{-1}$, a density filling factor
for clumps in the wind of 0.1, and a terminal wind velocity of 500 km
s$^{-1}$ found from the 1D stellar-wind modeling in \cite{H01}.

The length scale derived here for H$\alpha$ is larger than in the
recent VLTI aperture-synthesis imaging at Br$\gamma$ (2.166~\micron),
which revealed a fan-shaped outflow extending to 6--8 mas (14--19 AU)
along the NW--SW direction \citep{W16}. Since we find that convolving
with a 12 mas Gaussian better reproduces the observed H$\alpha$
profile than 6--8 mas Gaussians, the H$\alpha$-emitting region in
$\eta$ Car's wind is more extended than for Br$\gamma$, as expected.

The location of H$\alpha$ emission also holds some implications for
reconciling the clues of wind asymmetry with the average mass-loss
rate derived from 1D radiative transfer models of the unresolved
spectrum.  A relatively fast and dense polar wind is indicated by
P~Cygni absorption profiles seen in H$\alpha$, as noted above
\citep{S03a}.  Meanwhile, the H$\alpha$ emission components show
little angle dependence in these same spectra that view the star from
mid-latitudes to the pole \citep{S03a}.  This may be explained because
the H$\alpha$ emission (integrated over the whole wind) arises at
relatively large radii where the asymmetry is washed out, well outside
the regions of the strongest asymmetry of the bipolar wind seen in
near-IR interferometric data \citep{vanboekel03}.  P~Cygni absorption,
on the other hand, traces a single line of sight through the wind and
can still reveal the angle dependence of speed and density
\citep{S03a}.  The faster and denser polar wind stands in contrast to
lower mass-loss rates inferred from models of the colliding wind X-ray
emission.  Such models generally yield mass-loss rate estimates for
the primary star of $5\times10^{-4}$ $M_\sun$ yr$^{-1}$ or less
\citep{parkin09}.  This is less than the average mass-loss rate from
1D models, whereas the polar wind implies a higher mass-loss rate.
The likely explanation for this is that the colliding wind emission
reflects the density of the primary star's wind at low latitudes near
the equator, since the plane of the binary orbit is aligned with the
equatorial plane of the rotating primary star and with the equatorial
plane of the Homunculus \citep{madura12}, and the colliding wind shock
remains at low latitudes within about 45{\arcdeg}.  If the primary
star's wind were intrinsically spherical, it would be difficult to
understand the simultaneous indication of a higher mass-loss rate of
10$^{-3}$ $M_\sun$ yr$^{-1}$, which seems to be confirmed by our
H$\alpha$ imaging, and the lower mass-loss rates near the equator
indicated by X-ray emission.  The most natural conclusion is that the
primary star's wind does actually have a strong latitude dependence,
even though it may be modified further by the colliding wind shock
cone.

Finally, we note that while the weaker observed P Cygni
  absorption in H and Fe lines compared to the 1D model might imply
  that the mass-loss rate is a few times lower than predicted
  \citep{H01}, the absorption depth may also be influenced by the hot
  companion star's radiation, which is not included in the models, or
  by clumping in stellar wind (e.g., \citealt{P14}).  Recently,
  \cite{M10} attributed a decrease in the strength of emission lines
  to a decreasing mass-loss rate over time. This would be expected to
  shrink the H$\alpha$-forming region over time and make the FWHM of
  H$\alpha$ similar to that of 643 nm.  Our observations of the
  resolved H$\alpha$ profile at the present epoch appear to be
  consistent with the earlier mass-loss rate derived by \citet{H01} and do not confirm a reduction in the overall mass-loss rate since
  then.  Similarly, analysis of the continued monitoring of X-rays
  from the colliding wind emission in the $\eta$ Car system do not
  point to a reduction of the primary star's mass-loss rate over time
  \citep{R16,C17}.  Future diffraction-limited observations may be
  able to trace any possible variation in the mass-loss rate.

\section{Summary}
In 2016 and 2017, we used MagAO on the 6.5 m Clay telescope to perform
fast-exposure lucky imaging toward the massive evolved star
$\eta$~Carinae, using filters that sample H$\alpha$ (656 nm) and the
643 nm continuum. Our diffraction-limited images show that $\eta$ Car
is about 2.5--3 mas wider in H$\alpha$, which most likely comes from
the line-emitting region in its stellar wind. We show that this
H$\alpha$-forming region may have a size on the order of 12 mas, or
25--30 AU at the presumed distance of $\eta$~Car, in agreement with
the stellar-wind model of \cite{H01}.

\acknowledgements 
We thank the referee for helpful comments. This material is based upon work supported by the
National Science Foundation under grant No. 1506818 (PI Males) and NSF
AAG grant No. 1615408 (PI Close). Y.-L.W. and L.M.C. are supported by
the NASA Origins of Solar Systems award and the TRIF fellowship.
N.S.'s research on $\eta$ Carinae and related LBV-like eruptions received
support from NSF grant AST-1312221. K.M.M.'s and L.M.C.'s work is supported by the NASA Exoplanets Research Program (XRP) by cooperative agreement NNX16AD44G. This Letter includes data gathered with the 6.5 m Magellan Clay Telescope at Las Campanas Observatory, Chile.


\clearpage

\end{document}